\documentclass[a4paper,11pt]{article}
\usepackage{pos}

\title{Quark spin-orbit correlations in the proton}

\author*[a]{M.~Engelhardt}
\author[b]{J.~R.~Green}
\author[c]{N.~Hasan}
\author[d]{T.~Izubuchi}
\author[e]{C.~Kallidonis}
\author[f,g]{S.~Krieg}
\author[h]{S.~Liuti}
\author[i]{S.~Meinel}
\author[j]{J.~Negele}
\author[j]{A.~Pochinsky}
\author[k]{A.~Rajan}
\author[c,l]{G.~Silvi}
\author[m]{S.~Syritsyn}

\affiliation[a]{Department of Physics, New Mexico State University,
Las Cruces, NM 88003, USA}
\affiliation[b]{School of Mathematics and Hamilton Mathematics Institute,
Trinity College Dublin, Dublin 2, Ireland}
\affiliation[c]{Bergische Universit\"at Wuppertal, 42119 Wuppertal, Germany}
\affiliation[d]{Physics Department, Brookhaven National Laboratory,
Upton, NY 11973, USA}
\affiliation[e]{Thomas Jefferson National Accelerator Facility,
Newport News, VA 23606, USA}
\affiliation[f]{JARA \&\, IAS, J\"ulich Supercomputing Centre,
Forschungszentrum J\"ulich, 52425 J\"ulich, Germany}
\affiliation[g]{Helmholtz-Institut f\"ur Strahlen- und Kernphysik,
Universit\"at Bonn, 53115 Bonn, Germany}
\affiliation[h]{Department of Physics, University of Virginia,
Charlottesville, VA 22904, USA}
\affiliation[i]{Department of Physics, University of Arizona, Tucson,
AZ 85721, USA}
\affiliation[j]{Center for Theoretical Physics, Massachusetts Institute of
Technology, Cambridge, MA 02139, 
USA}
\affiliation[k]{Department of Physics, Old Dominion University,
Norfolk, VA 23529, USA}
\affiliation[l]{J\"ulich Supercomputing Centre,
Forschungszentrum J\"ulich, 52425 J\"ulich, Germany}
\affiliation[m]{Department of Physics and Astronomy, Stony Brook University,
Stony Brook, NY 11794, USA}

\emailAdd{engel@nmsu.edu}

\abstract{Generalized transverse momentum-dependent parton distributions
(GTMDs) provide a comprehensive framework for imaging the internal structure
of the proton. In particular, by encoding the simultaneous distribution of
quark transverse positions and momenta, they allow one to directly access
longitudinal quark orbital angular momentum, and, moreover, to
correlate it with the quark helicity. The relevant GTMD is evaluated
through a lattice calculation of a proton matrix element of a quark
bilocal operator (the separation in which is Fourier conjugate to the
quark momentum) featuring a momentum transfer (which is Fourier conjugate
to the quark position), as well as the Dirac structure appropriate for
capturing the quark helicity. The weighting by quark transverse position
requires a derivative with respect to momentum transfer, which is obtained
in unbiased fashion using a direct derivative method. The lattice
calculation is performed directly at the physical pion mass, using
domain wall fermions to mitigate operator mixing effects. Both the
Jaffe-Manohar as well as the Ji quark spin-orbit correlations are
extracted, yielding evidence for a strong quark spin-orbit coupling
in the proton.}

\FullConference{%
The 38th International Symposium on Lattice Field Theory, LATTICE2021\\
26th-30th July, 2021\\
Zoom/Gather@Massachusetts Institute of Technology
}

\begin{document}
\maketitle

\section{Introduction}
The dynamics of the angular momenta of the confined quarks in the proton
constitute a central topic of hadron structure physics. Their description
requires consideration of the full set of degrees of freedom available to
quarks; orbital angular momentum (OAM) depends both on quark position and
momentum, and the full quark angular momentum moreover includes the
quark spin. A combined accounting for these characteristics is provided
by (polarized) Wigner distributions that simultaneously encode position
and momentum, or, equivalently, generalized transverse momentum-dependent
parton distributions (GTMDs), which are related to the former by Fourier
transformation: The quark impact parameter, i.e., transverse position,
is Fourier conjugate to the transverse momentum transfer in the GTMD.
The GTMD framework moreover provides control over the ambiguity in
partitioning OAM among the quark and gluon degrees of freedom that is
inherent in a gauge theory. Owing to gauge invariance, quark degrees of
freedom cannot be considered in complete isolation, but must be accompanied
by gluonic admixtures. The two most prominent schemes for defining these
admixtures are the ones due to Ji \cite{jidecomp} and to Jaffe and Manohar
\cite{jmdecomp}. These schemes can be incorporated into the definition of
GTMDs in well-defined fashion.

The GTMD framework has been employed to directly evaluate quark OAM in
the proton in Lattice QCD calculations \cite{f14pap,f14deriv}, extending
the scope of lattice considerations of OAM beyond the traditional avenue
invoking Ji's sum rule \cite{cyprus}, which is focused specifically on the
Ji decomposition of OAM. A continuous, gauge-invariant interpolation between
Ji and Jaffe-Manohar quark OAM was obtained, revealing that Jaffe-Manohar
quark OAM is significantly enhanced in magnitude compared to Ji OAM, by
approximately 30\%. Upon incorporating methodological improvements of
the treatment of the momentum transfer, using a direct derivative method
\cite{f14deriv}, the result for Ji OAM was reconciled with the one obtained
Ji's sum rule, validating the approach.

Quark OAM itself does not reference the quark spin, and is obtained from
a GTMD in which the quarks are unpolarized. On the other hand, in view
of the strong chromodynamic fields through which a quark in the proton
propagates, the coupling of quark spin and OAM is expected to constitute
an important dynamical determinant of the angular momentum budget in the
proton. To obtain enhanced insight into the interplay between quark spin
and OAM, it is useful to quantify quark spin-orbit correlations in the
proton. These can be obtained in a manner that largely parallels the
calculation of quark OAM; essentially, the counting of quarks has to be
weighted by their spin, leading to the evaluation of a GTMD in which the
quarks are polarized (in the nomenclature of \cite{mms}, it is the GTMD
$G_{11} $ that encodes quark spin-orbit correlations). Here, a first
lattice evaluation of quark spin-orbit correlations in the proton is
presented, performed directly at the physical pion mass, employing a
domain wall fermion ensemble furnished by the RBC/UKQCD collaboration.

\section{Quark spin-orbit correlations}
The longitudinal quark spin-orbit correlation $\langle 2 L_3 S_3 \rangle $
in an unpolarized proton propagating in the 3-direction can be obtained
from a GTMD matrix element \cite{lorce},
\begin{equation}
\langle 2 L_3 S_3 \rangle = \frac{1}{2P^+ }
\epsilon_{ij} \frac{\partial }{\partial z_{T,i} }
\left. \frac{\partial }{\partial \Delta_{T,j} }
\frac{\langle p^{\prime } |
\overline{\psi} (-z/2) \gamma^+ \gamma^{5} U[-z/2,z/2] \psi(z/2)
| p \rangle }{ {\cal S} [U]}
\right|_{z^+ = z^- =0\, , \ \Delta_{T} =0\, , \ z_T \rightarrow 0}
\label{ldersingle}
\end{equation}
which differs from the matrix element that determines longitudinal quark
OAM, $\langle L_3 \rangle $, merely by the inclusion of the $\gamma^{5} $
structure weighting by helicity, and the fact that the proton state
needs to be longitudinally polarized if one wishes to access OAM itself.
A number of remarks are in order regarding eq.~(\ref{ldersingle}). The
in- and outgoing proton momenta $p,p^{\prime } $ differ by a transverse
momentum transfer $\Delta_{T} $, i.e., $p=P-\Delta_{T} /2$,
$p^{\prime } =P+\Delta_{T} /2 $, with the spatial component of $P$
pointing in 3-direction. Given that $\Delta_{T} $ is Fourier conjugate
to the quark impact parameter, $b_T $, taking the derivative with respect
to $\Delta_{T} $ and evaluating it at $\Delta_{T} =0$ corresponds to
taking the average of $b_T $. Conversely, the operator separation $z$ in the
quark bilocal operator, which is also chosen to be purely transverse,
$z\equiv z_T $, is Fourier conjugate to the transverse momentum $k_T $
of the quark. Thus, taking the derivative with respect to $z_T $ and
evaluating it at $z_T =0$ corresponds to taking the average of $k_T $.
Here, caution must be exercised with respect to the limit
$z_T \rightarrow 0$, which is associated with ultraviolet divergences.
In (\ref{ldersingle}), the $\Delta_{T} $ and $z_T $ dependences are
combined such as to yield the average $b_T \times k_T $, i.e.,
longitudinal OAM, correlated with the quark helicity through the
$\gamma^{+} \gamma^{5} $ Dirac structure. In view of the specification
$z^+ = z^- =0$, also the longitudinal quark momenta are integrated over.

\begin{figure}[h]
\begin{center}
\includegraphics[width=7cm]{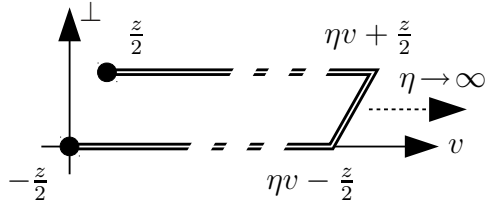}
\end{center}
\caption{Gauge connection path in the matrix element (\ref{ldersingle}).}
\label{figstaple}
\end{figure}

In addition to this kinematic structure, eq.~(\ref{ldersingle}) depends on
the choice of the Wilson line $U$, required by gauge invariance, that
connects the quark operators. It is accompanied by a combined multiplicative
soft and renormalization factor ${\cal S} [U] $ that compensates for the
divergences associated with $U$. As in previous lattice TMD and GTMD
studies \cite{tmdlat,bmlat,rentmd,f14pap,f14deriv},
this factor will be canceled by forming an appropriate
ratio of matrix elements in the further development below; as a result,
it does not need to be specified further for present purposes. The
choice of path for $U$ is the feature that allows one to account for
different definitions of quark OAM. As displayed in Fig.~\ref{figstaple},
staple-shaped paths $U\equiv U[-z/2,\eta v-z/2,\eta v+z/2,z/2]$
will be considered here, where the arguments of $U$ denote positions
that are joined by straight Wilson lines. The direction of the staple
is given by the vector $v$, and its length by $\eta $.

The special case $\eta =0$ corresponds to a straight Wilson line connecting
the quark operators directly. This case yields the quark OAM according to
the Ji decomposition of angular momentum \cite{jist}. On the other hand,
a staple-shaped path extending to infinity, $\eta \rightarrow \pm \infty $,
yields Jaffe-Manohar quark OAM \cite{hatta}. In the context of standard
TMDs, such a staple-shaped link is interpreted as incorporating final
state interactions of a struck quark in a (semi-inclusive) deep inelastic
scattering process as it is leaving the proton; the staple legs correspond
to hard, eikonal propagators of the struck quarks. Accordingly, the
difference between Jaffe-Manohar and Ji OAM has been interpreted
in terms of the torque experienced by a struck quark \cite{burk}. The
staple length $\eta $ of course can be varied (quasi-)continuously in a
lattice calculation, and one thus obtains a gauge-invariant, continuous
interpolation between the two limits.

Given the aforementioned physical role of the gauge link, a natural
direction $v$ for the staple would appear to be a lightlike vector,
along which the struck quark propagates away from the hadron remnant.
This choice, however, leads to rapidity divergences that call for
regularization, with a standard scheme being to rotate $v$ into the
spacelike region \cite{aybat,collbook}. Then, (\ref{ldersingle}) depends
on an additional Collins-Soper type rapidity regulator that can expressed
in Lorentz-invariant fashion,
\begin{equation}
\hat{\zeta } = \frac{v\cdot P}{\sqrt{|v^2 |} \sqrt{P^2 } }
\end{equation}
in terms of which the light-cone limit is approached for
$\hat{\zeta } \rightarrow \infty $. As will be described below, this
scheme is at the same time well-suited for casting the calculation of
(\ref{ldersingle}) as a lattice calculation.

It has already been noted above that, ultimately, a ratio of matrix
elements will be constructed that serves to cancel the combined
multiplicative soft and renormalization factor ${\cal S} [U] $.
A quantity suitable for this purpose in the present context is the
number of valence quarks
\begin{equation}
n = \frac{1}{2P^{+} }
\left. \frac{\langle p^{\prime } |
\overline{\psi}(-z/2) \gamma^+ U[-z/2,z/2] \psi(z/2)
| p \rangle }{ {\cal S} [U]}
\right|_{z^+ =z^- =0\, , \ \Delta_{T} =0\, , \ z_T \rightarrow 0}
\label{ndenom}
\end{equation}
which differs from (\ref{ldersingle}) in that the weighting with
$b_T \times k_T $ is omitted (thus, the matrix element simply counts
quarks), and also the helicity weighting with $\gamma^{5} $ is missing.
The latter choice is justified by the fact that domain wall fermions
are used in the present calculation, respecting chiral symmetry, and
therefore the vector and axial renormalization constants in the local
limits of the operators coincide. For discretizations that break chiral
symmetry, it would instead be appropriate to normalize specifically by
the axial charge rather than the vector one. The factor ${\cal S} [U] $
is even in $z_T $ and therefore cancels in the ratio
$\langle 2L_3 S_3 \rangle /n$.

Preserving chiral symmetry in the fermion discretization moreover
prevents the appearance of operator mixing effects
\cite{constmixtmd,shanamixtmd,greenmixtmd,jimixtmd} induced by the
breaking of chiral symmetry. Such effects would invalidate
the cancellation of renormalization factors through taking the
ratio $\langle 2L_3 S_3 \rangle /n$, since they would generate
additional additive terms in the numerator and denominator of the ratio.

At finite lattice spacing $a$, the derivative with respect to $z_T $ in
(\ref{ldersingle}) is realized as a finite difference. The renormalized
quantity therefore evaluated in practice is
\begin{equation}
\frac{\langle 2L_3 S_3 \rangle }{n} = \frac{1}{a} \epsilon_{ij}
\left. \frac{\frac{\partial }{\partial \Delta_{T,j} }
\left( \Phi^{[\gamma^{+} \gamma^{5} ]} (a\vec{e}_{i} )
- \Phi^{[\gamma^{+} \gamma^{5} ]} (-a\vec{e}_{i} ) \right) }{
\Phi^{[\gamma^{+} ]} (a\vec{e}_{i} )
+ \Phi^{[\gamma^{+} ]} (-a\vec{e}_{i} )}
\right|_{z^+ = z^- =0\, , \ \Delta_{T} =0}
\label{rdiscrete}
\end{equation}
where the transverse indices $i$ and $j$ are summed over, and
$\Phi^{[\Gamma ]} (z_T) = \langle p^{\prime } |
\overline{\psi}(-z/2) \Gamma U \psi(z/2) | p \rangle $. In the local limit
$z_T \rightarrow 0$, (\ref{rdiscrete}) contains additional divergences,
which, in view of the form of (\ref{rdiscrete}), are regularized by
cutting off transverse momenta at the overall ultraviolet resolution
of the calculation. This scheme a priori does not coincide with the
standard $\overline{MS} $ scheme, and a matching factor would be needed
to connect to the latter. The results of the quark OAM calculation
reported in \cite{f14deriv} suggest that this matching
factor, at least in the $\eta =0$ limit, does not deviate significantly
from unity within the statistical uncertainties typically achieved
in these GTMD calculations.

The derivative with respect to momentum transfer in (\ref{rdiscrete}), on
the other hand, is realized employing an unbiased direct derivative method
\cite{rome,nhasan}, as laid out in detail for the GTMD calculations at hand
in \cite{f14deriv}.

\section{Lattice calculation and results}
To cast the evaluation of (\ref{rdiscrete}) as a lattice calculation,
it is necessary to boost (\ref{rdiscrete}) to a Lorentz frame in which
the operator defining $\Phi^{[\Gamma ]} (z_T )$ exists at a single time.
This is the point for which a rapidity regulator scheme that renders
the staple direction $v$ spacelike, as described above, is crucial;
once all separations in the problem, i.e., $z_T $ and $\eta v$, are
spacelike, the requisite boost can be performed. After the boost,
$v$ points into the (longitudinal) 3-direction, and both $z_T $ and
$\Delta_{T} $ lie in the transverse plane, orthogonal to each other.

The lattice calculation was performed using a domain wall fermion
ensemble furnished by the RBC/UKQCD collaboration. The ensemble consisted
of 130 lattices of extent $48^3 \times 96$ with spacing $0.114\, \mbox{fm} $
and pion mass $m_{\pi } =139\, \mbox{MeV} $. An all-mode averaging scheme
with 33280 low-accuracy and 520 exact samples for bias correction was
employed. For this exploratory calculation,
the fairly small source-sink separation $8a=0.91\, \mbox{fm} $ was used
in order to control statistical fluctuations. Calculations were performed
for two proton momenta, $P_3 = 2\pi /(aL) $ and $P_3 = 4\pi /(aL) $ (where
$L=48$ is the spatial extent of the lattice). These correspond to the
rapidity regulator values $\hat{\zeta } =0.23$ and $\hat{\zeta } =0.46$.

\begin{figure}
\begin{center}
\includegraphics[width=10cm]{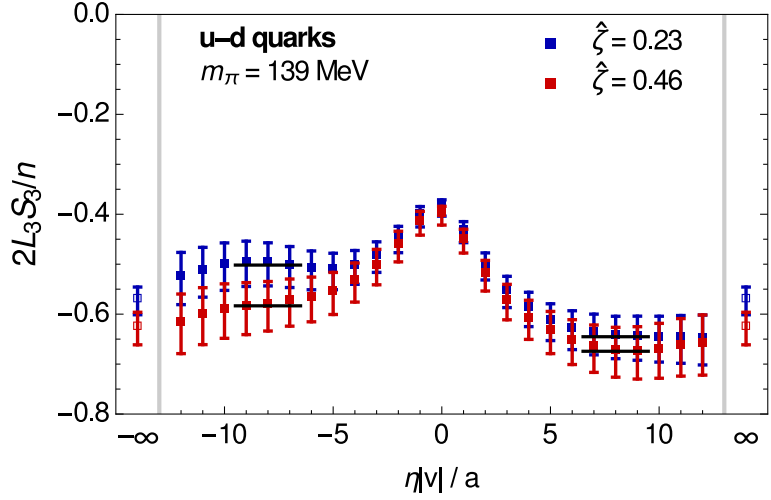}
\end{center}
\caption{Quark spin-orbit correlation in the proton as a function of
staple length $\eta $. The large-$|\eta |$ behavior is extracted from
the plateau fits indicated by the black lines, where the values displayed
for $\eta \rightarrow \pm \infty $ are obtained by averaging the plateaus
obtained on the positive and negative $\eta $ sides.}
\label{figso}
\end{figure}

The numerical results are summarized in Fig.~\ref{figso}. Data for the
isovector, $u-d$ quark combination are displayed, in which disconnected
diagram contributions to the proton matrix elements cancel; such disconnected
contributions were not evaluated in the present calculation. The quark
spin-orbit correlation $\langle 2 L_3 S_3 \rangle $ is shown as a function
of the staple length $\eta $, interpolating between the $\eta =0$ Ji
limit and the $\eta \rightarrow \pm \infty $ Jaffe-Manohar limit, which
is obtained by extrapolation. A marked dependence on $\eta $ is observed,
which is even stronger than in OAM, $\langle L_3 \rangle $, considered by
itself. In the latter case, the Jaffe-Manohar limit is enhanced in
magnitude compared to the Ji limit by about 30\% \cite{f14deriv}, whereas
for the spin-orbit correlation, the enhancement is approximately 50\%.

Comparing the results obtained for the two different values of the
Collins-Soper type rapidity regulator $\hat{\zeta } $, no strong dependence
can be discerned. Indeed, in the $\eta =0$ limit, there can be no dependence
on $\hat{\zeta } $, since in that limit there is no staple direction
$v$ on which the results could depend. However, even in the
large-$|\eta |$ Jaffe-Manohar limit, which in principle depends on
$\hat{\zeta } $, no statistically significant variation is evident between
the two surveyed values of $\hat{\zeta } $. This behavior was also seen
for OAM itself \cite{f14deriv}, where merely the $\hat{\zeta } =0$ data
available in that calculation deviate significantly from the non-zero
$\hat{\zeta } $ results. Seemingly, the large-$\hat{\zeta } $ limit is
approached rather quickly in these observables.

The spin-orbit correlation is negative and its magnitude is considerable.
As a point of reference, consider the uncorrelated product
$2 \langle L_3 \rangle \langle S_3 \rangle $ obtained in a longitudinally
polarized proton. Of course, taking separate unpolarized averages over proton
states, both factors in this product would be zero, whereas the spin-orbit
correlation $\langle 2 L_3 S_3 \rangle $ does not depend on whether the
proton is longitudinally polarized or the unpolarized average is taken. The
comparison is between, on the one hand, the indirect relative bias
between quark OAM and quark spin induced by the proton being in any
definite polarization state and, on the other hand, the direct correlation
between quark OAM and quark spin. Taking recourse to the recent comprehensive
study \cite{cyprus} of the Ji decomposition of proton spin at the physical
pion mass, one has, for the two light quark flavors,
\begin{eqnarray}
\langle L^u_3 \rangle = -0.22(3) \ , \ \
2\langle S^u_3 \rangle = 0.86(2) & \ \ \ \Rightarrow \ \ \ &
2\langle L^u_3 \rangle \langle S^u_3 \rangle = -0.19(3) \\
\langle L^d_3 \rangle = 0.26(2) \ , \ \
2\langle S^d_3 \rangle = -0.42(2) & \ \ \ \Rightarrow \ \ \ &
2\langle L^d_3 \rangle \langle S^d_3 \rangle = -0.11(1)
\end{eqnarray}
and, therefore, for the $u-d$ quark combination,
\begin{equation}
2\langle L^u_3 \rangle \langle S^u_3 \rangle -
2\langle L^d_3 \rangle \langle S^d_3 \rangle = -0.08(3)
\end{equation}
amounting to only $1/5$ of the direct correlation displayed in
Fig.~\ref{figso} at $\eta =0$. Therefore, there is a strong direct
dynamical coupling between quark OAM and spin, resulting in a correlation
that far exceeds the simple bias for these quantities induced by the proton
being in any definite polarization state. The quark OAM and spin are
(anti-)aligned rather rigidly, while the angular momentum of the quark
as a whole is, in comparison, less constrained in its orientation with
respect to the angular momentum of the rest of the constituents of the
proton. This is reminiscent of the $jj$ coupling scheme in atomic physics,
as opposed to the Russell-Saunders coupling scheme.

The result for the spin-orbit correlation obtained here is furthermore
consistent with a phenomenological estimate obtained in \cite{lorceso}
by connecting the spin-orbit correlation to moments of generalized
parton distributions (GPDs), cf.~also \cite{eomlirl} . For the $u-d$
quark combination, the estimate obtained in \cite{lorceso} in the
$\eta =0$ Ji limit, $\langle 2 L_3 S_3 \rangle \approx -0.37 $, agrees
well with the result exhibited in Fig.~\ref{figso}.

\section{Summary}
Generalized transverse momentum-dependent parton distributions (GTMDs)
provide a comprehensive framework for parametrizing the internal structure
of hadrons. The present work expands the scope of Lattice QCD calculations
of GTMD observables from the initial application to quark OAM in the proton
\cite{f14pap,f14deriv} to quark spin-orbit correlations, thus providing more
detailed insight into the dynamical determinants of the proton's internal
angular momentum structure. A strong, negative spin-orbit correlation is
found, which moreover is significantly enhanced in magnitude, by about 50\%,
when one transitions from the Ji to the Jaffe-Manohar definition of
OAM. By performing the lattice calculation using domain wall fermions
directly at the physical pion mass, systematic uncertainties that would
be engendered by breaking chiral symmetry in the fermion discretization,
as well as ones associated with chiral extrapolation were eliminated;
nonetheless, other systematic effects remain to be brought under control
in future work, among them, e.g., excited state contaminations that may
be significant for the fairly small source-sink separation employed in
the present study.

\acknowledgments
Computing time granted by the John von Neumann Institute for
Computing (NIC) and provided on the supercomputer JURECA \cite{jureca}
(Booster module) at J\"ulich Supercomputing Centre (JSC) is gratefully
acknowledged, as are resources provided by the U.S.~DOE Office of Science
through the National Energy Research Scientific Computing Center (NERSC),
a DOE Office of Science User Facility, under Contract No.~DE-AC02-05CH11231.
Calculations were performed employing the Qlua \cite{qlua} software suite.
S.M.~is supported by the U.S.~DOE, Office of Science, Office of High
Energy Physics under Award Number DE-SC0009913.
M.E., S.L., J.N.~and A.P.~are supported by the U.S.~DOE, Office
of Science, Office of Nuclear Physics through grants numbered
DE-FG02-96ER40965, DE-SC0016286, DE-SC-0011090 and DE-FC02-06ER41444,
respectively. S.S.~is supported by the National Science Foundation under
CAREER Award PHY-1847893. This work was furthermore supported by the
U.S.~DOE through the TMD Topical Collaboration.


\begin{thebibliography}{99}
\bibitem{jidecomp} X.~Ji, Phys. Rev. Lett. {\bf 78}, 610 (1997).
\bibitem{jmdecomp} R.~Jaffe and A.~Manohar, Nucl. Phys. {\bf B337},
509 (1990).
\bibitem{f14pap} M.~Engelhardt, Phys. Rev. {\bf D 95}, 094505 (2017).
\bibitem{f14deriv} M.~Engelhardt, J.~R.~Green, N.~Hasan, S.~Krieg, S.~Meinel,
J.~Negele, A.~Pochinsky and S.~Syritsyn,
Phys. Rev. {\bf D 102}, 074505 (2020).
\bibitem{cyprus} C.~Alexandrou, S.~Bacchio, M.~Constantinou, J.~Finkenrath,
K.~Hadjiyiannakou, K.~Jansen, G.~Koutsou, H.~Panagopoulos and G.~Spanoudes,
Phys. Rev. {\bf D 101}, 094513 (2020).
\bibitem{mms} S.~Mei\ss ner, A.~Metz and M.~Schlegel,
JHEP {\bf 0908}, 056 (2009).
\bibitem{lorce} C.~Lorc\'e and B.~Pasquini,
Phys. Rev. {\bf D 84}, 014015 (2011).
\bibitem{tmdlat} B.~Musch, P.~H\"agler, M.~Engelhardt, J.~Negele
and A.~Sch\"afer, Phys. Rev. {\bf D 85}, 094510 (2012).
\bibitem{bmlat} M.~Engelhardt, P.~H\"agler, B.~Musch, J.~Negele
and A.~Sch\"afer, Phys. Rev. {\bf D 93}, 054501 (2016).
\bibitem{rentmd} B.~Yoon, M.~Engelhardt, R.~Gupta, T.~Bhattacharya,
J.~R.~Green, B.~Musch, J.~Negele, A.~Pochinsky, A.~Sch\"afer, and
S.~Syritsyn, Phys. Rev. {\bf D 96}, 094508 (2017).
\bibitem{jist} X.~Ji, X.~Xiong and F.~Yuan,
Phys. Rev. Lett. {\bf 109}, 152005 (2012).
\bibitem{hatta} Y.~Hatta, Phys. Lett. {\bf B708}, 186 (2012).
\bibitem{burk} M.~Burkardt, Phys. Rev. {\bf D 88}, 014014 (2013).
\bibitem{aybat} S.~M.~Aybat and T.~Rogers,
Phys. Rev. {\bf D 83}, 114042 (2011).
\bibitem{collbook} J.~C.~Collins, {\it Foundations of Perturbative QCD}.
Cambridge University Press, 2011.
\bibitem{constmixtmd} M.~Constantinou, H.~Panagopoulos and G.~Spanoudes,
Phys. Rev. {\bf D 99} (2019) 074508.
\bibitem{shanamixtmd} P.~Shanahan, M.~Wagman and Y.~Zhao,
Phys. Rev. {\bf D 101} (2020) 074505.
\bibitem{greenmixtmd} J.~Green, K.~Jansen and F.~Steffens,
Phys. Rev. {\bf D 101} (2020) 074509.
\bibitem{jimixtmd} Y.~Ji, J.-H.~Zhang, S.~Zhao and R.~Zhu,
Phys. Rev. {\bf D 104} (2021) 094510.
\bibitem{rome} G.~M.~de~Divitiis, R.~Petronzio and N.~Tantalo,
Phys. Lett. {\bf B718}, 589 (2012).
\bibitem{nhasan} N.~Hasan, J.~R.~Green, S.~Meinel, M.~Engelhardt, S.~Krieg,
J.~Negele, A.~Pochinsky and S.~Syritsyn,
Phys. Rev. {\bf D 97}, 034504 (2018).
\bibitem{lorceso} C.~Lorc\'e, Phys. Lett. {\bf B735}, 344 (2014).
\bibitem{eomlirl} A.~Rajan, M.~Engelhardt and S.~Liuti,
Phys. Rev. {\bf D 98}, 074022 (2018).
\bibitem{jureca} J\"ulich Supercomputing Centre, {\it JURECA: Modular
supercomputer at J\"ulich Supercomputing Centre}.
Journal of Large-Scale Research Facilities {\bf 4}, A132 (2018).
http://dx.doi.org/10.17815/jlsrf-4-121-1
\bibitem{qlua} A.~Pochinsky, {\it Qlua}. https://usqcd.lns.mit.edu/qlua.
\end{thebibliography}
\end{document}